\documentstyle[mnextra]{mn}
\input{epsf}
\seceqnum           
\def\etal{{\rm et al.} }
\def\kms  {{\rm km \, s^{-1}}}
\def\Mpc  {{\it h}^{-1}\, {\rm Mpc}}

\def\Msol {{\it h}^{-1}\, {\rm M_\odot}}
\def\Lsol {{\it h}^{-2}\, {\rm L_\odot}}
\def\d    {{\rm d}}
\def\ie {{\rm i.e. }}
\def\eg {{\rm e.g. }}
\def\log {{\rm log}}

\def\lsim{\mathrel{\hbox{\rlap{\hbox{\lower4pt\hbox{$\sim$}}}\hbox{$<$}}}}
\def\gsim{\mathrel{\hbox{\rlap{\hbox{\lower4pt\hbox{$\sim$}}}\hbox{$>$}}}}

\def\bj{b_{\rm J}}

\begin{document}
\title[2PIGG abundances]
{Galaxy groups in the 2dFGRS: the number density of groups}
\author[V.R. Eke et al.]
{\parbox[t]\textwidth{
V.R. Eke$^1$,
Carlton M. Baugh$^1$,
Shaun Cole$^1$,
Carlos S. Frenk$^1$,
Julio F. Navarro$^{2,}\thanks{CIAR and Guggenheim Fellow}$}
\vspace*{6pt} \\
{$^1$Department of Physics, University of Durham, South Road,
    Durham DH1 3LE, UK} \\
$^2$Department of Physics and Astronomy, University of Victoria, Victoria,
    BC, V8P 5C2, Canada}
\maketitle

\begin{abstract}

The abundance of galaxy clusters as a function of mass is
determined using the 2dFGRS Percolation-Inferred Galaxy Group (2PIGG)
catalogue. This is used to estimate the amplitude of the
matter fluctuation spectrum, parametrised by the linear theory
rms density fluctuations in spheres of $8 \Mpc$, $\sigma_8$. The
best-fitting value for this parameter is highly correlated with the
mean matter density in the Universe, $\Omega_{\rm m}$, and is found to satisfy
$\sigma_8=0.25~\Omega_{\rm m}^{-0.92-4.5(\Omega_{\rm m}-0.22)^2} \pm
10\%$ (statistical) $\pm 20\%$ (systematic) for
$0.18\le\Omega_{\rm m}\le0.50$, assuming that 
$\Omega_{\rm m}+\Omega_\Lambda=1$. This gives $\sigma_8=0.89$ when
$\Omega_{\rm m}=0.25$. A $\sim 20$ per cent correction has been
applied to undo the systematic bias inherent in the measurement
procedure. Mock catalogues, constructed from large cosmological N-body
simulations, are used to help understand and model these
systematic errors. The abundance of galaxy groups as a function of
group $\bj$ band luminosity is also determined. This
is used in conjunction with the halo mass function, determined from
simulations, to infer the variation of halo mass-to-light ratio over
four orders of magnitude in halo mass. The mass-to-light ratio shows a
minimum value of $100 hM_\odot/L_\odot$ in the $\bj$ band at a total
group luminosity of $L_{b_{\rm J}} \approx 5\times10^9\Lsol$. Together
with the observed Tully-Fisher relation, this implies that
the observed rotation speed of Tully-Fisher galaxies is within $\sim
10$ per cent of the typical circular speed of haloes hosting brightest
galaxies of the same luminosity.

\end{abstract}

\section{Introduction}

The abundance of galaxy clusters provides a very direct way to estimate
$\sigma_8$, the linear theory $rms$ density fluctuations in comoving spheres of
$8\Mpc$, extrapolated to redshift zero (\eg Peebles, Daly \& Juszkiewicz
1989; Frenk \etal 1990; White, Efstathiou \& Frenk 1993; Eke, Cole \&
Frenk 1996; Viana \& Liddle 1996; Ikebe \etal 2002; Pierpaoli \etal
2003; Schuecker \etal 2003; Henry 2004). This parameter sets the 
normalisation of the matter power spectrum. As the cluster abundance varies
rapidly with $\sigma_8$, the main difficulty in this method is not the
determination of the abundance, but knowing the mass of the objects under
consideration. The current $\sim 10$ per cent uncertainty on the
value of $\sigma_8$ (see Viana \etal 2003 and Henry 2004 for recent
discussions of cluster-based estimates) brackets important
differences in the 
formation histories of dark matter haloes. This has implications for
a broad range of topics, such as the redshift at which
gravitationally bound structures first form, the merger histories of
galaxy-sized haloes and the concentration of the dark matter density
profile within haloes. 

Many of the cluster samples used so far to
determine $\sigma_8$ have been based on the ROSAT X-ray All Sky Survey.
It is clearly desirable to supplement these estimates with
others derived from independent cluster samples, where the clusters
are found using different techniques. There are already a number of
studies that have estimated the galaxy cluster mass function using
catalogues of optically selected objects (\eg Bahcall \& Cen 1993;
Biviano \etal 1993; Girardi \etal 1998;
Mart\'inez \etal 2002; Bahcall \etal 2003). In this study, the
2dFGRS Percolation-Inferred Galaxy Group (2PIGG) catalogue (Eke \etal 2004a)
is used, in conjunction with mock galaxy catalogues, to provide a new
estimate of the mass function. The use of mock catalogues
extends the previous work by quantifying some of the systematic biases
inherent in the method.

The group luminosity function, that is the abundance of groups as a
function of their total luminosity, has previously been determined
using galaxy redshift survey data by Moore, Frenk \& White (1993),
Marinoni, Hudson \& Giuricin (2002) and Mart\'inez \etal (2002). The
2PIGG catalogue increases
the number of available groups by a factor of approximately two
relative to the work of Mart\'inez \etal (2002), who used the earlier
data release from the 2dFGRS. The group luminosity function provides
an indirect way to measure the mass function if one knows how to
relate mass to 
luminosity. For instance, if the statistical uncertainties on the
cluster dynamical mass estimates are significantly larger than those
on the cluster luminosities, then the combination of the group
luminosity function with a typical cluster mass-to-light ratio will
yield a more accurate estimation of the cluster mass function than
would be found from directly using the dynamical masses. Turning this
around, given a theoretical mass function and a measured group luminosity
function, one can infer the mapping from mass to luminosity as a
function of group size (Marinoni \& Hudson 2002). This picks out the
group scales at which mass is most efficiently turned into starlight
and, as such, encodes important clues about the process of
galaxy formation (Benson \etal 2000; van den Bosch, Yang \& Mo 2003).

In this paper, mock galaxy catalogues, constructed from N-body
simulations of the $\Lambda$CDM cosmology combined with a
semi-analytical model for placing galaxies into dark matter haloes,
are used to estimate how 
group-finding in the 2dFGRS leads to biases in the recovered cluster
mass and group luminosity functions. Then the 2PIGG catalogue is used
to estimate the cluster mass and group luminosity functions, and the
mass-to-light ratio for groups. This final quantity is inferred under
the assumption that the group mass function is given by the fit to the
halo mass function measured in N-body
simulations, as described by Jenkins \etal (2001, J01). Given the typical halo
mass, or equivalently circular speed, at a given luminosity, one can
compare these results with the observed Tully-Fisher
relation (Tully \& Fisher 1977; Bell \& de Jong 2001). Under the
assumption that observed Tully-Fisher galaxies lie in typical haloes,
this determines the conversion from observed rotation speeds of
galaxies to halo circular velocities. This result has some bearing on the
long-running debate concerning the ability of galaxy formation models
to match the galaxy luminosity function and the Tully-Fisher relation
simultaneously (Kauffmann, White \& Guiderdoni 1993; Cole \etal 1994;
Heyl \etal 1995).

Section~\ref{sec:data}
contains a brief description of the mock and real catalogues 
used in this study. The mass function calculation and the constraint
this imposes upon $\sigma_8$ is described in
Section~\ref{sec:mf}. Section~\ref{sec:lf} contains the group luminosity
function results, which are applied to determining the group
mass-to-light ratio variation in Section~\ref{sec:mol}. In
Section~\ref{sec:tf}, the halo mass-to-light ratios are compared with
the observed Tully-Fisher relation.

\section{Brief description of the mock and real catalogues}\label{sec:data}

The construction of the 2PIGG catalogue is described in detail by Eke
\etal (2004a). This paper also discusses the generation of mock galaxy
catalogues from a combination of dark matter N-body
simulations and semi-analytical galaxy formation models. The catalogue
has been constructed from the two contiguous patches in the 2dFGRS
(Colless \etal 2001), which contain a total of $\sim 190\,000$
galaxies. Of these, about $55$ per cent are placed into $\sim 29\,000$
groups containing at least two members. The median redshift of the
groups is $0.11$, like that of the galaxies, and the reliability of
the estimated group masses and luminosities has been gauged using 
mock catalogues as described by Eke \etal (2004b).
\begin{figure}
\centering
\centerline{\epsfxsize=8.5cm \epsfbox{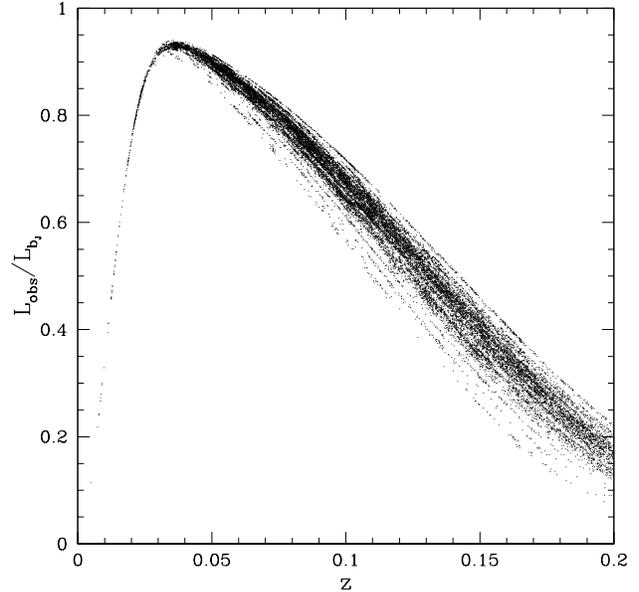}}
\caption{The ratio of observed to total group luminosity for each mock
group as a function of redshift. The total group luminosity is
estimated using the global galaxy luminosity Schechter function
described in the text to correct for galaxies outside the 2dFGRS flux limits.}
\label{fig:correctl}
\end{figure}

Mocks have been created from N-body simulations of the standard
$\Lambda$CDM model with parameter values $\Omega_{\rm m}=0.3$,
$\Omega_\Lambda=0.7$ and
$\sigma_8=0.90$ and $0.71$. The $\sigma_8=0.90$ simulation is the
$\Lambda$CDM2 run described by Jenkins \etal (1998), whereas the lower
$\sigma_8$ simulation cube contains $288^3$ particles in a box of length
$154 \Mpc$, corresponding to a very similar mass resolution.
Both the semi-analytical models of Cole \etal
(2000) and Benson \etal (2003) have been used to place galaxies
into the dark matter distributions in the two simulation cubes. Note
that the luminosities of the 
semi-analytical mock galaxies have been scaled by small,
colour-preserving amounts so that the mock galaxy $\bj$-band
luminosity function exactly matches that found in the 2dFGRS.
Dynamical masses of groups are estimated using
\begin{equation}
M=A\frac{\sigma^2 r}{G},
\label{mass}
\end{equation}
where $A=5$ (Eke \etal 2004a), 
$\sigma$ is the 1-dimensional velocity dispersion,
calculated using the gapper algorithm (Beers, Flynn \& Gebhardt 1990)
and removing $85\,\kms$ in quadrature to account for redshift
measurement errors, and $r$ is the {\it r.m.s.} projected separation of
galaxies from the group centre, assuming an $\Omega_{\rm m}=0.3,
\Omega_\Lambda=0.7$ cosmological model. Observed group luminosities are
estimated by summing the luminosities of the individual galaxies with
their associated weights to account for spectroscopic incompleteness:
\begin{equation}
L_{\rm obs}=\Sigma_i^{N} w_i L_{i}.
\label{easyw}
\end{equation}
This observed group luminosity is then corrected for galaxies lying
outside the survey flux limits assuming a global
Schechter galaxy luminosity function
\begin{equation}
\phi(L)\d L=\phi_* \left(\frac{L}{L_*}\right)^\alpha {\rm exp}\left({-\frac{L}{L_*}}\right)
\frac{\d L}{L_*},
\label{sch}
\end{equation}
with $(M_*-5\log_{10}h,\alpha)=(-19.725,-1.18)$ (Eke \etal 2004b) and
$M_\odot=5.33$ in the $\bj$ band. Fig.~\ref{fig:correctl} shows the
ratio of $L_{\rm obs}$ to the total group luminosity, $L_{b_{\rm J}}$.
Note that there are upper and lower flux
limits imposed on the galaxies in the 2dFGRS, hence the shape traced
out by the galaxy groups in Fig.~\ref{fig:correctl}, with the upper
flux limit rejecting bright galaxies in the nearby groups, and the
lower flux limit being important at higher redshifts. The stripes
traced out by the points are a result of the distinct flux limits in
different patches of the survey.

In all of what follows, the group sample will be restricted to the
$z<0.12$ objects, because at higher redshifts, less than half of the
total group luminosity is contained in visible galaxies and the
correction for this missing light becomes too uncertain (see
Fig.~\ref{fig:correctl}). For the same reason, the volume at $z<0.02$
will also be excised. 

When calculating the abundance of groups, it is necessary to know
the volume surveyed. The maximum volume within which a particular group
could have been detected by a survey is $V_{\rm max}$. If a group
is defined to comprise at least $N$ galaxies, then the group
$V_{\rm max}$ could be approximated by the $V_{\rm max}$ of the $N$th most
luminous galaxy in that group. This takes into account the variable
sky coverage as a function of flux in the 2dFGRS. However, for $N>1$,
this ignores the possibility that the set of galaxies linked together
by the friends-of-friends (FOF) algorithm may not remain grouped at different
redshifts, when the linking volume has changed in size and the set of
galaxies satisfying the flux limits of the survey may have changed to
the extent that parts of the group are no longer joined. To deal with
these issues exactly would be awkward. These complications are minor
though, because the linking volume changes with redshift to take into
account the varying sampling of the galaxy distribution. 
Consequently, the $V_{\rm max}$ of the $N$th
most luminous galaxy is used to define the group $V_{\rm max}$
throughout the rest of this paper. The mock catalogues can be used to
gauge the extent to which this choice biases the results.
Of course, for the group luminosity function, $N_{\rm min}=1$ can be
used, in which case the $V_{\rm max}$ value is correct.

\section{Cluster mass functions}\label{sec:mf}

\begin{figure}
\centering
\centerline{\epsfxsize=8.5cm \epsfbox{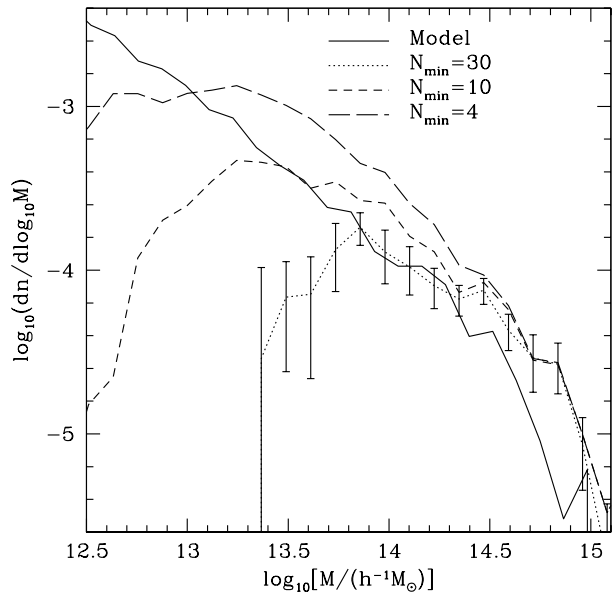}}
\caption{Model and mock recovered mass functions. The solid line traces
the mass function of dark matter haloes in the simulation cube (the
$\Lambda$CDM2 run by Jenkins \etal 1998). The
other lines show the mass function inferred using different minimum
numbers of galaxies to define the group sample, as indicated in the
key. Statistical errors, shown by the error bars, are
calculated using the scatter between $22$ mock catalogues constructed
from the Hubble Volume simulation.}
\label{fig:mfbias}
\end{figure}

\begin{figure}
\centering
\centerline{\epsfxsize=8.5cm \epsfbox{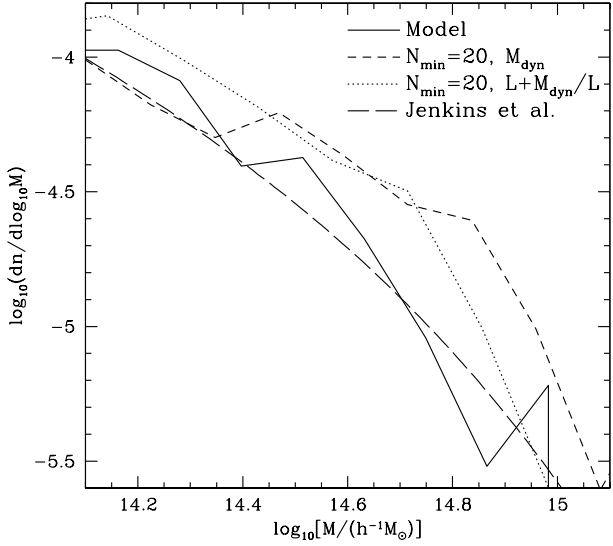}}
\caption{Mass functions for dark haloes in the simulation and for
groups identified in the mock catalogue. A solid line traces
the mass function of dark matter haloes in the simulation cube. Also
plotted are mass functions for mock groups inferred using
dynamical masses (short-dashed line) and group luminosities with a
shift for the typical mass-to-light ratio (dotted
line). The long-dashed line represents the J01
fitting formula to describe the dark matter halo mass 
function appropriate for the power spectrum of the model.}
\label{fig:mfbias2}
\end{figure}

\begin{figure}
\centering
\centerline{\epsfxsize=8.5cm \epsfbox{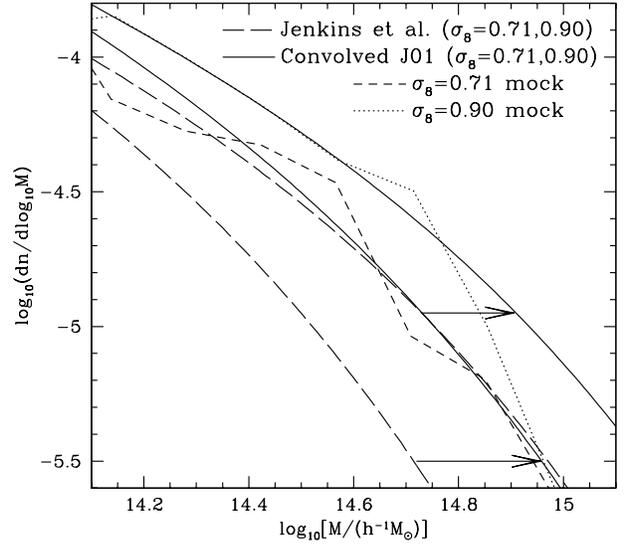}}
\caption{The $\sigma_8$ dependence of the recovered mass
function. Long-dashed lines show the J01 fitting function
describing the mass function for the two different values of
$\sigma_8$ used to make mock catalogues. The solid lines show the
effect of measurement errors on these curves, as described
in the text. A dotted line traces the mass function
recovered from the $\sigma_8=0.9$ mock, whereas the short-dashed line
is the corresponding curve for $\sigma_8=0.71$. The
horizontal arrows show the impact of the measurement bias.}
\label{fig:mfbias3}
\end{figure}

The accuracy with which dynamical masses can be estimated varies
rapidly with the number of galaxies sampling the group potential
(see figure 3 of Eke \etal 2004b). Errors in estimated masses
coupled with the steep decline in the abundance of clusters, \ie
groups with masses above $\sim 10^{14}\Msol$, as a
function of increasing mass, lead to systematic biases in the inferred
abundance of clusters. This happens because more low-mass
objects are scattered to 
higher masses than are scattered down from those higher masses.
Consequently, the abundance of high mass galaxy clusters will be
systematically overestimated as a result of the uncertainties in
measuring the individual cluster masses.
This effect is shown in Fig.~\ref{fig:mfbias} for the mock catalogues,
which compares the mass functions recovered from groups samples of different
$N_{\rm min}$ at $z\le0.12$ in the mocks with the 
mass function of dark matter haloes in the simulation cube. The model
mass function shows the abundance of dark matter haloes found using a
friends-of-friends (FOF) group-finder with a linking length $b=0.2$
times the mean interparticle separation (Davis \etal 1985). Error bars
represent the 
standard deviation among the results from $22$ different mock
catalogues made using the Hubble Volume simulation (Evrard \etal
2002). Norberg \etal (2002) provide a detailed description of how the
mock catalogues are constructed. Hawkins \etal (2003) have shown that
the galaxies in mock 
catalogues made from the Hubble Volume simulation cluster like those
in the 2dFGRS, suggesting that the use of these mocks is appropriate.
Cosmic variance only increases the statistical
errors by $\sim 10$ per cent relative to the $1/V_{\rm max}$-weighted
Poisson errors.
Note that restricting the sample to include only groups containing
large numbers of galaxies introduces incompleteness at
lower masses, which, to some extent, counteracts the overestimation of
the mass function due to mass measurement errors. Nevertheless,
as Fig.~\ref{fig:mfbias} shows, the abundance is still overestimated by
a factor of $\sim 2$ for $M\gsim3\times10^{14}\Msol$. For the
remainder of this Section, $z_{\rm max}=0.12$ and $N_{\rm min}=20$
will be used. These cuts restrict the sample to $\sim 350$ clusters.

The dynamically-inferred group masses show more scatter about the
true halo mass in the simulations than
masses inferred using the total group luminosity and a typical
mass-to-light ratio (Eke \etal 2004b). Thus, one can produce a more
robust estimate of the cluster mass function by using the cluster
total luminosity function and a typical mass-to-light ratio. This is
shown, at least for the highest masses, in Fig.~\ref{fig:mfbias2}. The
median mass-to-light ratio for all mock groups with
$L>10^{11.5}\Lsol$ ($\Upsilon=471~h M_\odot/L_\odot$) has been used as a global
shift to the cluster luminosity function to infer the mass function (dotted
line). The mass-to-light ratio varies little for these most massive
objects (Eke \etal 2004b), so this simple shift is appropriate.
Also shown is the J01 fitting formula for the
mass function that describes the population of haloes from which the
model curve is sampled. Hereafter in this paper, only recovered mass
functions obtained using the cluster luminosity function
and the cluster mass-to-light ratio, as shown by the dotted
line in Fig.~\ref{fig:mfbias2}, will be considered.

In order to place constraints upon the normalisation of the power
spectrum, it is necessary firstly to model the bias in recovering the mass
function and secondly to compare the measured mass function in the
2PIGG catalogue with the
biased models for each value of $\sigma_8$. Fig.~\ref{fig:mfbias3}
shows the J01 mass functions for two different values
of $\sigma_8$, before and after applying a shift to account for the
measurement errors. An approximation to the bias introduced by
measuring the mass function is to translate the J01 curves
to the right by $0.17$ and $0.20$ for $\sigma_8=0.90$ and $0.71$
respectively. These shifts provide the best $\chi^2$ fits (assuming
weighted Poisson errors) to the mass functions
measured from the mock group catalogues for both values of $\sigma_8$. The
uncertainty on the size of these translations, judging by the change
in $\chi^2$ of the resulting fits, is $\sim 7$ per cent in mass for
the Cole \etal (2000) mocks. The shifts required for the mocks
made using the Benson \etal (2003) semi-analytical procedure differ by $\sim
7$ per cent from the Cole \etal (2000) mocks. Thus, one should consider these
corrections to be uncertain by $\sim 10$ per cent in mass.
With these shifts, each model curve can
be converted into a mass function in measurement space.
Note how the measurement bias is larger for the
lower value of $\sigma_8$, where the clusters are more difficult to
detect because of the lower contrast. This, unfortunately, reduces the
power of the test, but it is nevertheless apparent that these two
different values of $\sigma_8$ can still be distinguished.

\begin{figure}
\centering
\centerline{\epsfxsize=8.5cm \epsfbox{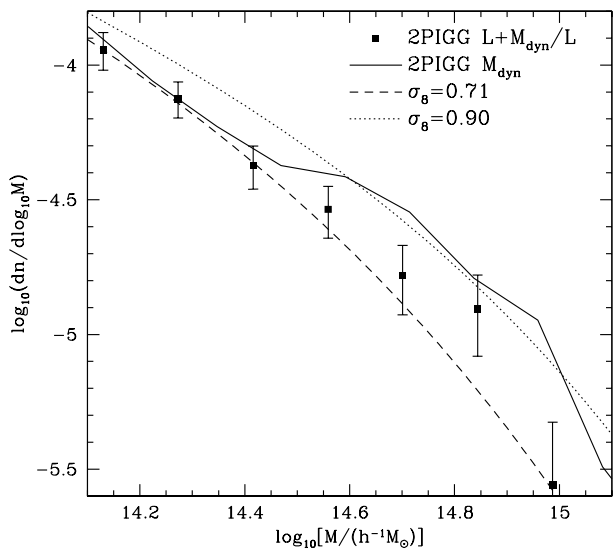}}
\caption{2PIGG and model group mass functions. The model mass
functions after convolution with the measurement errors, shown with
solid lines in the previous figure, are now shown 
with dotted ($\sigma_8=0.90$) and dashed
($\sigma_8=0.71$) lines. The points with error bars show the 2PIGG
results using the cluster luminosity function and typical
mass-to-light ratio. A solid line shows the 2PIGG mass function based
on the dynamical mass measurements of equation~\ref{mass}.}
\label{fig:2piggmf}
\end{figure}

\begin{figure}
\centering
\centerline{\epsfxsize=8.5cm \epsfbox{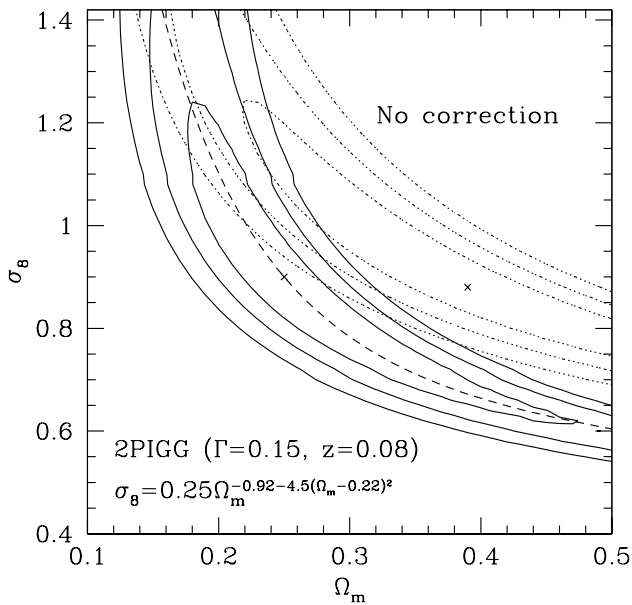}}
\caption{The constraints on $\sigma_8$ and $\Omega_{\rm m}$ from the
2PIGG catalogue. The $\chi^2$ contours are shown, with
(solid lines) and without (dotted lines) the correction for the bias
in the recovery of the mass function. The contours contain $68$, $95$
and $99.7$ per cent of probability, and a cross marks the best-fitting
parameter set for each case. The dashed line shows a fit to the most
probable $\sigma_8(\Omega_{\rm m})$ for the bias-corrected contours.}
\label{fig:rawb}
\end{figure}

\begin{figure}
\centering
\centerline{\epsfxsize=8.5cm \epsfbox{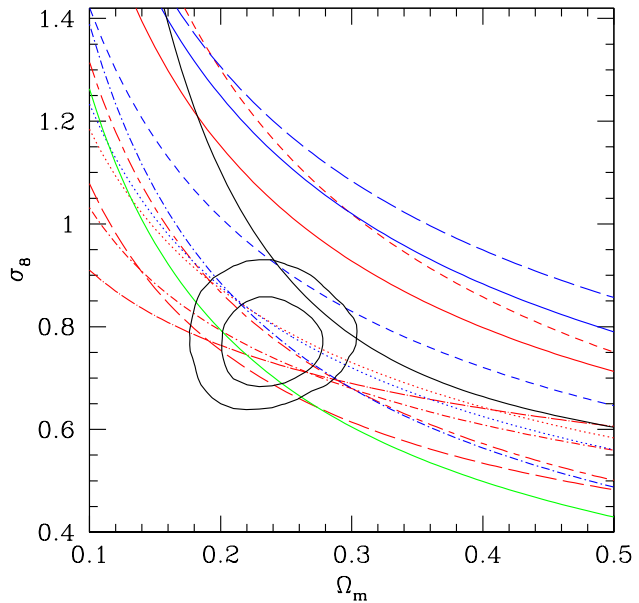}}
\caption{Comparison of the 2PIGG results (solid black line) with
various other estimates of the amplitude of matter fluctuations. The
red lines show measurements based largely on the cluster abundance:
Eke, Cole \& Frenk (1996, solid line); Pierpaoli \etal (2001, short
dashed); Bahcall \etal (2003, short-long dashed); Reiprich \&
B\"ohringer (2002, dot-short dashed); Seljak (2002, dotted); Viana,
Nichol \& Liddle (2002, long dashed); Allen \etal (2003, dot-long dashed). Blue
lines trace the amplitudes inferred using weak lensing-based
estimates from: Brown \etal (2003, dotted); Rhodes \etal (2004, long dashed);
Heymans \etal (2005, dot-short dashed); Massey \etal (2005, solid);
van Waerbeke, Mellier \& Hoekstra (2005, short dashed). The green line
shows the results of Tinker \etal (in preparation), who use a model of
the halo occupation distribution, in conjunction with the galaxy
correlation functions for different luminosity galaxies in the 2dFGRS
and the 2PIGG mass-to-light ratio variation with group size. The
rounded 68 and 95 per cent probability contours are from figure 12 of
Sanchez \etal (2005), and result from a joint analysis of the power
spectra of both the CMB and 2dFGRS. 
}
\label{fig:compb}
\end{figure}

The mass functions inferred from the 2PIGG data are shown in
Fig.~\ref{fig:2piggmf}. Once again, the directly determined mass
function, based on dynamical masses, gives higher abundances at this mass
than the mass function recovered from the luminosity function and the typical
cluster mass-to-light ratio ($\Upsilon=429~h M_\odot/L_\odot$). Note
that the 2PIGG results generally lie between the convolved model
curves for $\sigma_8=0.71$ and $0.90$ ($\Omega_{\rm m}=0.3$ in both cases).
Error bars again are the $1/V_{\rm max}$-weighted Poisson errors.
This neglects the $\sim 10$ per cent
additional contribution from cosmic variance, which is minor relative
to the systematic uncertainties inherent in this method.

By assuming that the convolving function generates a mean shift that varies
linearly with $\sigma_8$ and not at all with $\Omega_{\rm m}$, one can
extrapolate the convolved model to a range of $\Omega_{\rm m}$ and
$\sigma_8$, in order to determine the best-fitting values of these
parameters. The results of this exercise are shown in
Fig.~\ref{fig:rawb}. Only the 5 data points with
$14.4\le\log[M/(\Msol)]\le15$ are used in the $\chi^2$ fit. The contours
contain $68$, $95$ and 
$99.7$ per cent of the probability. Dotted and solid lines show the
constraints on the parameters before and after the correction for the
measurement bias is applied. In short, this correction changes the
best-fitting parameters from $(\Omega_{\rm m}=0.39,\sigma_8=0.88)$ to
$(0.25,0.90)$. The dashed line showing the variation of the
best-fitting $\sigma_8$ as a function of $\Omega_{\rm m}$ in the range
$0.18\le\Omega_{\rm m}\le0.50$ has the equation:
\begin{equation}
\sigma_8=0.25\Omega_{\rm m}^{-0.92-4.5(\Omega_{\rm m}-0.22)^2}.
\label{sig8fit}
\end{equation}
Note that these contours have all been calculated under the assumption
that the CDM transfer function has a shape given by $\Gamma=0.15$
(Bardeen \etal 1986; Sugiyama 1995) as has recently been measured
using the 2dFGRS (Cole \etal 2005). Using values of $\Gamma=0.1$ or
$0.25$ changes the value of $\sigma_8$ by less than $5$ per cent for 
$\Omega_{\rm m}=0.3$. While the statistical uncertainty upon the
estimate of $\sigma_8(\Omega_{\rm m})$ is small ($\sim 10$ per
cent), the systematic correction that has been applied to account for
the measurement bias is substantially bigger ($\sim 20$ per cent). It
is worthwhile reiterating that the probability contours involve
extrapolating the noisy fits of convolving functions to the mock
recovered mass functions, so this systematic shift is quite rough. As
was discussed above, the correction is uncertain by
$\sim 10$ per cent in mass. This corresponds roughly to $5$ per cent
systematic uncertainty in the estimated $\sigma_8$ value from this
shift. There is another systematic uncertainty associated with the
calibration of the mass estimation (the value of $A$ in equation~\ref{mass}).

An illustration of the relative importance of systematic uncertainties
in current estimates of $\sigma_8$ is contained in
Fig.~\ref{fig:compb}, which shows the results from a number of
different studies. The 2PIGG line is shown in black, along with
other cluster abundance-based estimates in red, and weak lensing-based
results in blue. The black probability contours show the results of
Sanchez \etal (2005), who jointly analysed the CMB and 2dFGRS power
spectra. The green line shows the best-fitting parameters inferred
using the 2dFGRS correlation functions, a model for the halo
occupation distribution and the 2PIGG mass-to-light ratios (Tinker
et al., private communication). Quoted errors
on the various curves are usually of the order of $10-15$ per
cent. Some of the variation from one study to another will come from
assuming different power spectrum shapes ($\Gamma$), but this
dependence is quite weak (Seljak 2002; Rhodes \etal 2004) so the excess
scatter between the curves is a sign of other systematic effects. For
the cluster abundance-based measurements, the main source of
systematic differences comes due to the choice of the mass-observable
relation (see Henry 2004).

\section{Group total luminosity function}\label{sec:lf}

\begin{figure}
\centering
\centerline{\epsfxsize=8.5cm \epsfbox{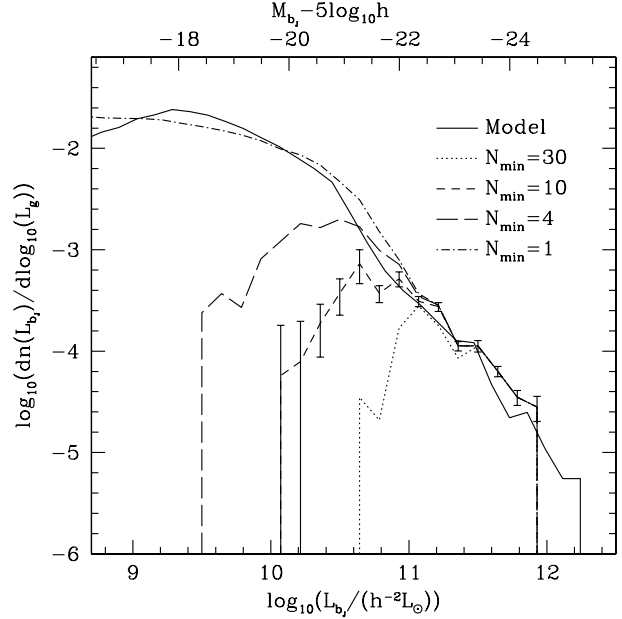}}
\caption{The abundance of groups as a function of their total
$\bj$-band luminosity. A solid line traces the actual group luminosity
function in the simulation, and the other lines show the recovered
functions using all mock
groups with $z<0.07$ and $N\ge N_{\rm min}$. Error bars are shown only
for the $N_{\rm min}=10$ case, and are calculated using the scatter
between $22$ different mocks made from the Hubble Volume simulation.}
\label{fig:dlf1}
\end{figure}

The total $\bj$-band group luminosity function is a distribution that,
unlike the galaxy luminosity function,
can readily be compared with theoretical dark matter halo mass
functions. In Section~\ref{sec:mol} the group luminosity function will
be used in conjunction with theoretical mass functions to derive halo
mass-to-light ratios.

Fig.~\ref{fig:dlf1} shows how well the $\bj$-band group
luminosity function can be recovered from a mock catalogue.
Only the groups with $0.02\le z\le0.07$ are included in the calculation of
these curves. As the upper limit in redshift is increased, the typical
correction for luminosity in galaxies fainter than the flux limit
increases. Consequently, the recovered luminosity functions become
increasingly biased high. On the other hand, if the maximum redshift
is too restrictive, then there are insufficient clusters, and the
volume probed may no longer be representative of the Universe as a
whole. The choice of $z_{\rm max}=0.07$ is a compromise 
between these two competing factors. Even with this value, the group
luminosity function recovered from the mock catalogue is still biased slightly
high relative to the true model function for the most luminous systems
in the simulation.
Note how the value of $N_{\rm min}$ impacts upon the abundance of 
low luminosity groups. The best recovery of the model occurs when
$N_{\rm min}=1$. Some depletion of the less luminous
groups is evident: these objects sometimes contaminate the
bigger groups, causing the abundance of the more luminous systems to
be slightly overestimated.

\begin{figure}
\centering
\centerline{\epsfxsize=8.5cm \epsfbox{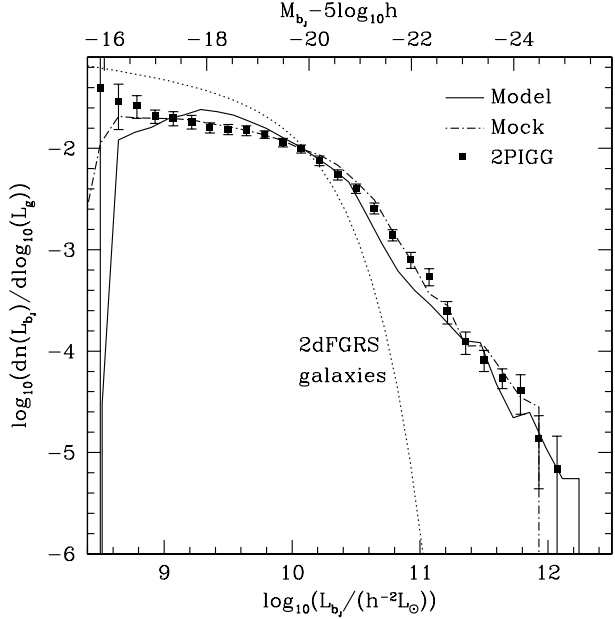}}
\caption{The abundance of groups as a function of their total
$\bj$-band luminosity. A solid line traces the model group luminosity
function, the dot-dashed line shows the recovered function using all mock
groups with $z<0.07$ and $N\ge 1$. The points show the group
luminosity function found using the 2PIGG catalogue.}
\label{fig:dlf2}
\end{figure}

Fig.~\ref{fig:dlf2} shows how the group luminosity function recovered
from the 2PIGG catalogue compares with that in a mock catalogue
constructed from a simulation with $\sigma_8=0.9$ using the Cole \etal
(2000) semi-analytical model. Only groups at $z\le0.07$ that contain
at least one detectable galaxy are included in the calculation. The
2PIGG results accurately trace those recovered from the mock.
Also shown is the 2dFGRS galaxy luminosity function, which
crosses the group luminosity function at $\sim L_*$, the
characteristic luminosity in the galaxy luminosity function (Norberg
\etal 2002). Many of
the fainter galaxies reside in groups with $L>L_*$; hence the
abundance of low luminosity groups lies beneath that of the galaxies.

\begin{figure}
\centering
\centerline{\epsfxsize=8.5cm \epsfbox{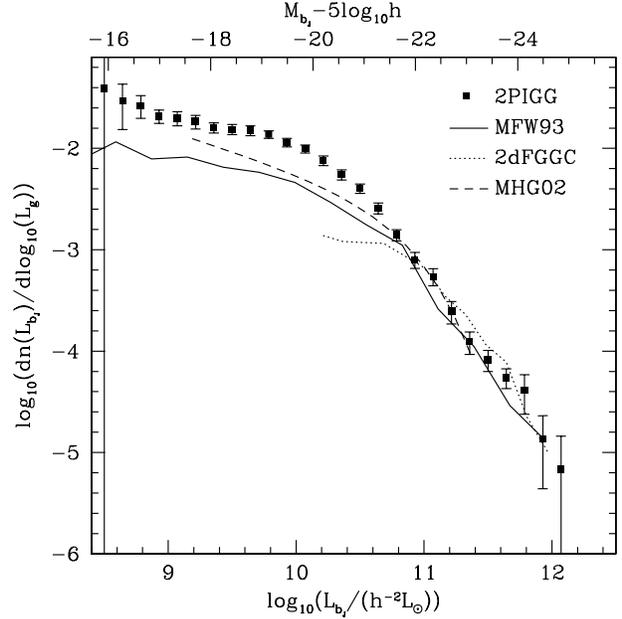}}
\caption{The abundance of groups as a function of their total
$\bj$-band luminosity. The points show the group luminosity function
found using the 2PIGG catalogue out to $z=0.07$, whereas the lines
show the results of Moore, Frenk \& White (1993, solid), Marinoni,
Hudson \& Giuricin (2002, dashed) and Mart\'inez \etal (2002, dotted).}
\label{fig:dlf3}
\end{figure}

Fig.~\ref{fig:dlf3} shows how the 2PIGG results compare with other
published work. The group luminosity functions of Marinoni \etal
(2002), from the Nearby Optical Galaxy (NOG) catalogue, Moore, Frenk
\& White (1993), from the CfA survey, and Mart\'inez \etal (2002),
using the 2dF Galaxy Group Catalogue (2dFGGC), are all quite similar
to that from the 2PIGGs for group 
luminosities above $\sim 10^{11}\Lsol$. At lower luminosities, the
2dFGGC results lie lower, because of the requirement that their groups
contain at least $4$ galaxies. The 2PIGG group luminosity function also
lies somewhat above those of Moore \etal and Marinoni et al. Note that
the Marinoni \etal curve has been shifted $0.55$ magnitudes fainter
from their published fit to
account for both the $0.3$ magnitudes difference between their $B$
band (RC3 asymptotic photometric system) and $\bj$ (Marinoni, private
communication), and $0.25$ 
magnitudes of internal absorption that they had included. The Moore
\etal curve has been shifted under the assumption that $\bj=B_{\rm
Zwicky}-0.05$. This should make these curves directly comparable
with the other results in the $\bj$ band. Both the Moore \etal and
Marinoni \etal studies probe
smaller volumes than in the 2PIGG case, so cosmic variance may be partly
responsible for the lower abundance of $\lsim 10^{11}\Lsol$ groups found in these earlier studies. Also, the group-finder
used by Marinoni \etal places a higher fraction of galaxies into
groups, so one might expect that fewer isolated galaxies would remain,
yielding fewer low luminosity groups.

\section{Group mass-to-light ratios}\label{sec:mol}

\begin{figure}
\centering
\centerline{\epsfxsize=8.5cm \epsfbox{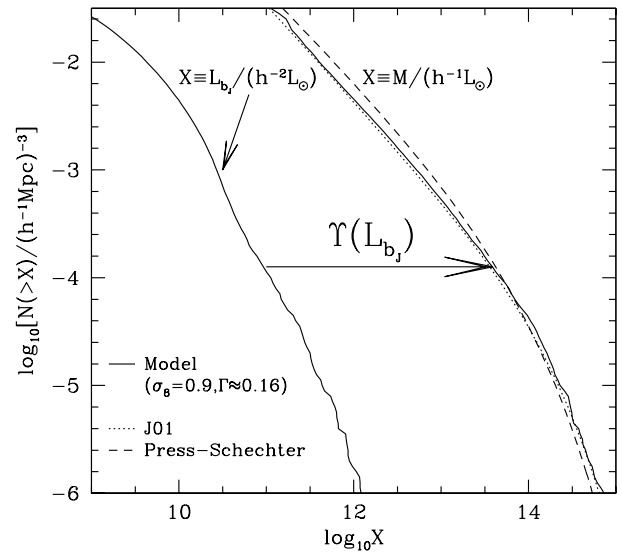}}
\caption{Cumulative group abundances as functions of both mass and
luminosity in the simulation model. The quantity $X$ on the horizontal
axis, is either the group $\bj$-band luminosity in $\Lsol$ or the
group mass in $\Msol$. A dotted line shows the J01
fit to the mass function, whereas the dashed line is the
corresponding Press-Schechter curve. The horizontal distance with an
arrow represents the inferred group mass-to-light ratio for groups
with a luminosity equal to $10^{11} \Lsol$.}
\label{fig:marhud}
\end{figure}

By comparing, at a fixed abundance, the measured cumulative group
luminosity function with a 
mass function motivated by numerical simulations of a CDM model, it
is possible to determine the mass-to-light ratio down to groups that
contain only a single visible galaxy in the 2dFGRS. This method (see
Marinoni \& Hudson 2002, MH02) is shown schematically in
Fig.~\ref{fig:marhud}, where the mass-to-light ratio gives the
mapping from the cumulative luminosity function to the cumulative mass
function. An assumption of this method is that the group mass varies
monotonically with luminosity, so that this mapping is
unique. Fig.~\ref{fig:marhud} shows the importance of the choice of mass
function. For instance, the Press \& Schechter (1974) formula
underestimates the abundance of the largest clusters and overestimates
that of Local Group-sized objects. This overabundance would produce
mass-to-light ratios for small groups that are too high by $\sim 50$
per cent.

This is an indirect, model-dependent measurement of the mass-to-light
ratio, because the mass 
function shape and amplitude are assumed to be of a particular form,
rather than being directly measured. However, with this method one can
probe to smaller group sizes, where group luminosities are still
well-defined but the masses are difficult to measure. The uncertainty
associated with the accuracy of
the adopted mass function includes a non-negligible
contribution from uncertainty in the value of $\sigma_8$ for the real
Universe. As discussed in Section~\ref{sec:mf}, the abundance of such 
clusters depends very sensitively upon $\sigma_8$. Consequently, the
mass-to-light ratios inferred from this method for these large
objects will suffer from this uncertainty. However, for smaller
groups, this effect is minor, and the mass function is quite robust to
changes in $\sigma_8$. Thus, this method for measuring the
mass-to-light ratio is complementary to the direct measurement
technique, which is most effective for the larger systems that have enough
galaxies to allow an accurate mass estimation.

\begin{figure}
\centering
\centerline{\epsfxsize=8.5cm \epsfbox{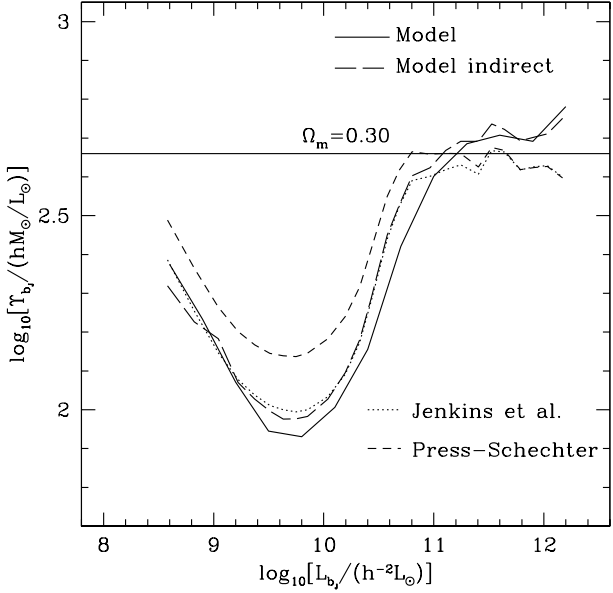}}
\caption{Model mass-to-light ratios as functions of group
luminosity. The solid curve traces the median mass-to-light ratio
in bins of group luminosity, whereas the long dashed line represents
the function required to map the model group luminosity function to
the model group mass function. If, instead of using the actual group
mass function from the simulation, the J01 or Press-Schechter
formulae are used, then the dotted and short dashed curves result.}
\label{fig:diffmol}
\end{figure}

It is worth noting that the mass-to-light ratio as a function of group
luminosity obtained by matching abundances is not the same as
the median mass-to-light ratio of
groups in each particular luminosity bin. This is because there
is some scatter in the group masses at each group luminosity. 
However, the difference between these two mass-to-light ratios in the
model can be seen in Fig.~\ref{fig:diffmol} to be small. Also
shown are the results of inferring the mass-to-light ratio assuming
the J01 and Press-Schechter mass functions rather than using
the actual mass function. This illustrates the $50$ per cent
overestimation for small haloes mentioned in the discussion of the
preceding figure. The J01 mass function is used throughout
the rest of this paper.

\begin{figure}
\centering
\centerline{\epsfxsize=8.5cm \epsfbox{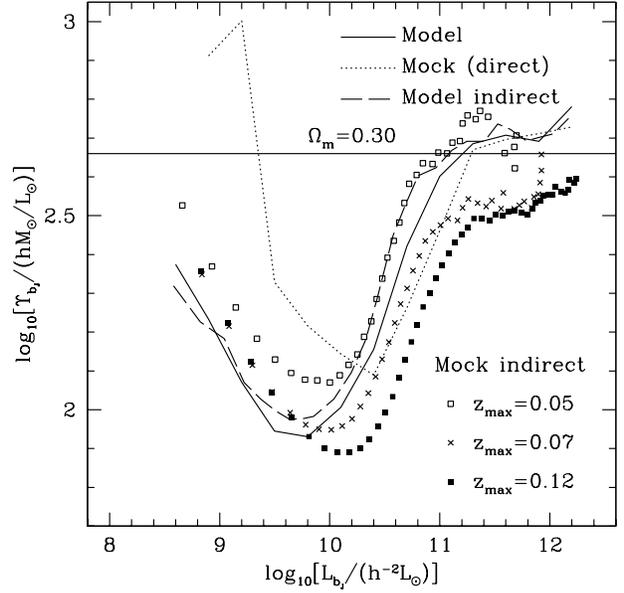}}
\caption{The recovery of the mass-to-light ratio as a function of group
luminosity. The dotted curve traces the median mass-to-light ratio
recovered from the mock catalogue using the 
dynamically inferred masses. This is the attempt to recover the model
represented by the solid curve. The three sets of points show the
mass-to-light ratio inferred from different sets of mock
groups using the abundance-matching technique, and are attempts to
measure the long-dashed curve of the model.}
\label{fig:molmh1}
\end{figure}

\subsection{Results}

\begin{figure}
\centering
\centerline{\epsfxsize=8.5cm \epsfbox{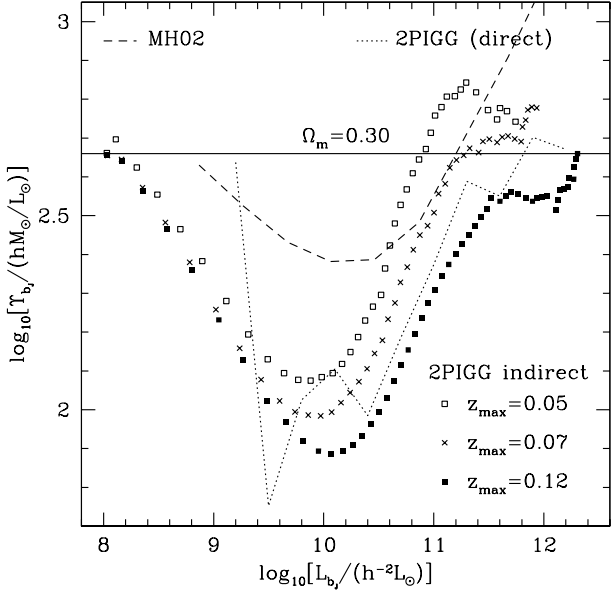}}
\caption{The recovery of the mass-to-light ratio as a function of group
luminosity for the 2PIGG sample. The dotted curve traces the median
mass-to-light ratio recovered using the 
dynamically inferred masses. The three sets of points show the
mass-to-light ratio variation inferred from different sets of 2PIGG
groups using the abundance-matching technique. A dashed line shows the
results of Marinoni \& Hudson (2002) from the NOG sample of galaxy groups.}
\label{fig:molmh2}
\end{figure}

The systematic errors in the recovery of the group luminosity function
alluded to in Section~\ref{sec:lf} impact significantly upon the
inferred variation of the group mass-to-light
ratio. Fig.~\ref{fig:molmh1} shows how changing the maximum redshift
of the mock group sample used to measure the luminosity function
affects the results. For the largest groups, the mass-to-light ratio
drops with increasing $z_{\rm max}$, as the abundance of clusters
becomes more overestimated because of increasing
contamination by interlopers and larger corrections for luminosity in
galaxies beneath the flux limit. For groups with $L\lsim
10^{10}\Lsol$ however, the volume probed is smaller and the errors
associated with the correction from observed to total group luminosity
are smaller. Consequently, the group luminosity functions  are less
dependent upon $z_{\rm max}$. The direct method for inferring the
group mass-to-light ratio becomes significantly biased for $L\lsim
10^{10}\Lsol$, so this group luminosity is an appropriate value at
which to switch between the direct and indirect methods. In this way, 
the `corrected' mass-to-light ratio can be recovered
over almost four orders of magnitude in group luminosity. For
$L>10^{10}\Lsol$ the 
appropriate correction factor is the difference between the directly
recovered curve (dotted) in Fig.~\ref{fig:molmh1} and the model curve
(solid). For $L<10^{10}\Lsol$ the correction factor is chosen to be the
difference between the crosses and the model shown by the solid curve,
so that the final `corrected' curve provides an estimate of the
typical mass-to-light ratio as a function of group luminosity.

The uncorrected results for the 2PIGG groups are shown in
Fig.~\ref{fig:molmh2} and compared with those of MH02.
As for the mock catalogues, the systematic differences between
the different $z_{\rm max}$ samples are evident for the bigger groups.
The increasing overestimation of the group luminosity function, as
$z_{\rm max}$ increases, gives rise to decreasing inferred
mass-to-light ratios. Note that the inferred mass-to-light ratio for
the smallest groups is comparable with that measured directly for the
clusters. 

It is apparent that the 2PIGG results are
somewhat different from those of MH02. This discrepancy arises for two
main reasons. For the small groups, MH02 use the Press-Schechter
formula for the mass function. As was seen in
Fig.~\ref{fig:molmh1}, this leads to an overestimation of the
mass-to-light ratio of small groups by $\sim 25$ per cent (this, of
course, assumes that the underlying mass function is that appropriate
for $\Lambda$CDM). This difference is compounded at $L_{b_{\rm
J}}<10^{10.6} \Lsol$ by the different group luminosity functions 
measured from the 2PIGG and NOG samples, as was
shown in Fig.~\ref{fig:dlf3}. As the NOG sample contains a
lower abundance of groups at $L\lsim10^{10.6} \Lsol$, the inferred
mass-to-light ratio is larger than for the 2PIGG sample. Note that
the MH02 results in Fig.~\ref{fig:molmh2} differ from those plotted
in figure $15$ by Eke \etal (2004b), because they did not use the
appropriate waveband correction.

\begin{figure}
\centering
\centerline{\epsfxsize=8.5cm \epsfbox{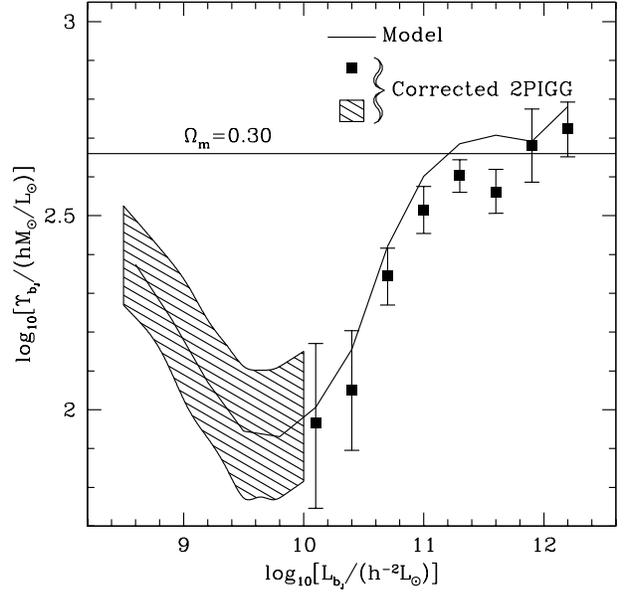}}
\caption{The `corrected' mass-to-light ratio as a function of group
luminosity for the 2PIGG sample. The error bars include the
statistical errors added in quadrature to the uncertainty in the
systematic shift 
applied to `correct' the measurements. Directly measured mass-to-light
ratios are shown with points, whereas those inferred assuming that the
global halo mass function is given by the fitting function of J01
are shown by the shaded region.}
\label{fig:molmh3}
\end{figure}

Having calibrated the biases introduced by the two methods to measure
group mass-to-light ratios with the mock catalogues, it is now
possible to `correct' the 2PIGG results in
Fig.~\ref{fig:molmh2}. Using the $z_{\rm max}=0.07$ data for the
indirect mass-to-light ratio measurement below $L=10^{10} \Lsol$
yields the results shown in Fig.~\ref{fig:molmh3}. The errors on the
indirectly measured points come from the scatter between $22$ mock
catalogues made from the Hubble Volume simulation, added in
quadrature to the uncertainty in the correction factor. For the $L>10^{10}
\Lsol$ results, the error bars show the statistical uncertainty in the
raw mass-to-light ratio added in quadrature to the uncertainty in the
correction factor. The uncertainties in the correction factors are
calculated by taking eight 
different mock catalogues with different $\sigma_8$ values and
semi-analytical schemes, and finding the scatter between the corrections.

The corrected mass-to-light ratio variation with luminosity for the 2PIGGs
shows a plateau at cluster masses, a decrease to a minimum at
$L\sim10^{10} \Lsol$, and an increase for smaller groups. The value of
the halo mass at this minimum is $\sim 10^{12}\Msol$, which agrees
well with the weak lensing analysis of Hoekstra, Yee \& Gladders
(2004). They found a halo of this luminosity to have
$M_{200}=(8.4\pm0.7\pm0.4)\times10^{11}\Msol$, which should be
increased by $\sim 20$ per cent to compare with the halo mass definition
used here. It is reassuring that these two completely different
methods show such consistency.

One might wonder how surprising it is that the corrected, recovered
mass-to-light variation looks so similar to that in the model for
$L<10^{10} \Lsol$, where the masses of the 2PIGGs are not directly
measured. After all, the model mass function has been assumed to be appropriate
for the real Universe, and the galaxy luminosity function in the model has
been scaled so as to match that of the 2dFGRS. The only
possibility for the corrected 2PIGG results to differ from the model
in Fig.~\ref{fig:molmh3} is if the group luminosity functions differ,
despite the identical global galaxy luminosity functions. Thus, the
measurement of an increase in 2PIGG mass-to-light ratio as group size
decreases from $L\sim10^{10} \Lsol$ is, if not inevitable, to be
expected merely from the fact that the faint end slope of
the galaxy luminosity function is flatter than that of the
$\Lambda$CDM mass function at low masses. The extra information
available here comes from the fact that the group luminosity function
is tracing the same structures probed with the mass function
measurements. Consequently, the normalisation of the mass-to-light 
ratio is not predetermined but, in fact, recovered in addition to the
variation with group size.

\section{The Tully-Fisher relation}\label{sec:tf}

\begin{figure}
\centering
\centerline{\epsfxsize=8.5cm \epsfbox{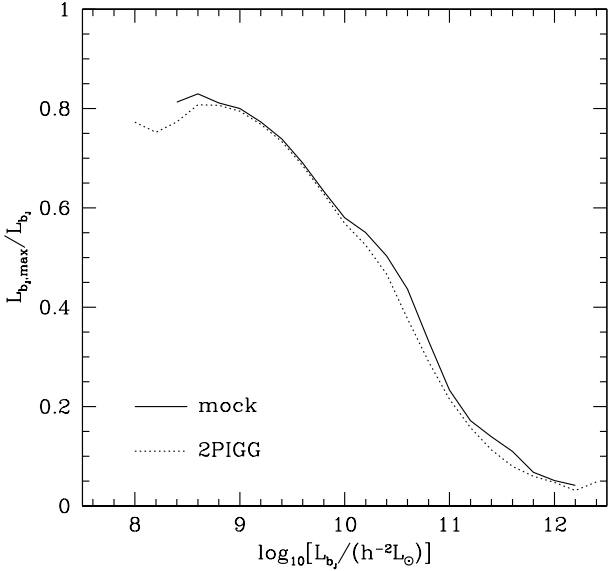}}
\caption{The ratio of the median brightest galaxy luminosity to the
total group luminosity as a function of luminosity for the mock groups
and 2PIGGs, shown with solid and dotted lines respectively.}
\label{fig:tf2}
\end{figure}

\begin{figure}
\centering
\centerline{\epsfxsize=8.5cm \epsfbox{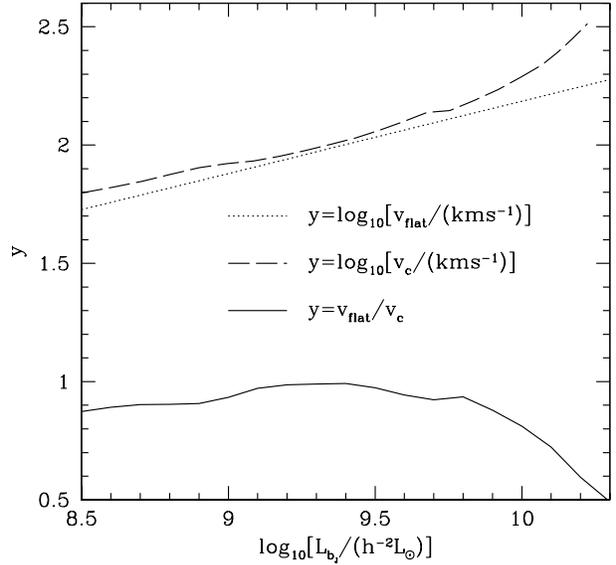}}
\caption{The galaxy rotation speed, $v_{\rm flat}$, and inferred
halo circular speed, $v_{\rm c}$, as functions of blue
luminosity. For the galaxy rotation speed curve (dotted line), the
luminosity is that of the galaxy, 
according to the Tully-Fisher relation of Bell and de Jong (2001). The
luminosity for the dashed curve is the total group luminosity
multiplied by the ratio shown in Fig.~\ref{fig:tf2} to convert it to
the typical luminosity of the brightest galaxy in each halo. The solid
line shows the ratio of these two velocities as a function of primary
galaxy luminosity.}
\label{fig:tf}
\end{figure}

The results in Fig.~\ref{fig:molmh3} show how the typical halo mass
varies with halo luminosity. For a particular definition of a halo,
one can convert this to a relation between the halo circular speed and
the halo luminosity via
\begin{equation}
L_{b_{\rm J}}=\left(\frac{3}{4\pi\Delta\rho_{\rm crit}}\right)
^\frac{1}{2}\frac{1}{\Upsilon G^\frac{3}{2}} ~v_{\rm c}^3,
\label{tf}
\end{equation}
where $\Delta\rho_{\rm crit}$ represents the mean enclosed density of
the halo, taken to be $100$ times the critical value. (Note that this
value is not exact because the identification of the
groups is done using a FOF algorithm, rather than growing a sphere out
to a particular density contrast.) This relation is reminiscent of the
Tully-Fisher relation between galaxy luminosity and 
rotation speed at the edge of the visible galaxy. For low luminosity
systems, a central bright galaxy usually dominates the total group
luminosity, so the galaxy and group luminosities are likely to be
quite similar. Fig.~\ref{fig:tf2} shows the median ratio of the
brightest galaxy to total group luminosity, as a function of total
group luminosity. The mock and 2PIGG curves are very similar, with the
main difference being that the brightest galaxy in the bigger mock
groups typically contains a slightly higher fraction of the total group
luminosity than in the 2PIGGs. 

If one were to apply the median shift of Fig.~\ref{fig:tf2}
to convert the group luminosity to the typical luminosity of the
brightest contained galaxy, then equation~(\ref{tf}) could relate
galaxy luminosity to halo circular speed at the virial radius. Then, one
might reasonably ask, given the Tully-Fisher relation and the results
in Fig.~\ref{fig:molmh3}, what is the relationship between the galaxy
rotation speed and the halo circular speed for galaxies with $L\lsim 10^{10}
\Lsol$? The answer to this question is contained in
Fig.~\ref{fig:tf}. The $B$-band Tully-Fisher relation of
Bell \& de Jong (2001) has been converted to $b_J$, and is shown with
a dotted line. Note that these authors advocate the use of $v_{\rm
flat}$ to characterise the velocity derived from the galaxy rotation curve.
The dashed curve shows the halo circular velocity from
equation~(\ref{tf}) as a function of luminosity (ie the group luminosity
multiplied by the factor shown in Fig.~\ref{fig:tf2} to convert it to
a galaxy luminosity). A solid line traces the
ratio of $v_{\rm flat}$ to $v_{\rm c}$.
How should this figure be interpreted? If one considers a
galaxy with $L_{b_{\rm J}}\sim10^{10}\Lsol$, then the Tully-Fisher relation
of Bell \& de Jong gives $v_{\rm flat}\sim
160\,\kms$, whereas the abundance matching method implies that it
typically lives in a halo with circular speed $\sim 200\,\kms$. For less
luminous galaxies, these two velocities become more similar, so that the
flat part of the galaxy rotation curve corresponds to a speed that is
approximately the same as the circular velocity of the host halo.

The suggestion that the galaxy rotation speed is similar to the halo
circular velocity is broadly in accord with the results for the
$I$-band Tully-Fisher relation in the semi-analytical galaxy formation
models of Cole \etal (2000) and Croton \etal 2005. They found that the
galaxy luminosity 
function and the Tully-Fisher relation could be matched simultaneously
when the galaxy rotation speed was assumed to equal the halo circular
velocity. However, one would expect as baryons concentrate in
the centres of galaxy-sized potential wells, they would drag in
dark matter, leading to galaxy rotation speeds that are a
few tens of per cent larger than the halo circular velocity. 
In this case, the Tully-Fisher relation is no longer reproduced by the
model. Similar conclusions have been drawn from hydrodynamical
simulations of galaxy formation by
Navarro \& Steinmetz (2000) and Eke, Navarro \& Steinmetz (2001), and
also from the analysis of de Jong \etal (2004). 

It is possible that the comparison above could be misleading; perhaps
the galaxies included in Tully-Fisher samples are atypical. If the
observed set of galaxies lie in either unusually luminous or unusually low
concentration haloes, then this might reconcile the apparent discrepancy
between the low observed rotation speeds and the higher expected
values. For example, the semi-analytical model behind the mock
catalogues used in this work predicts that the distribution of masses
of haloes with $L\sim10^{10}\Lsol$ has a $1\sigma$ scatter that
equates to $\sim 20$ per cent in halo circular speed. If, for some
reason, the galaxies selected in the
Tully-Fisher measurements lie in the less
massive haloes at that particular luminosity, then this could have a
very significant impact in reducing the dark matter contribution to
the observed rotation curves. Similarly, if typical
Tully-Fisher galaxies happen to lie in atypically unconcentrated
haloes, then the dark matter contribution to the observed rotation
curves would again be reduced.

\section{Conclusions}\label{sec:conc}

The 2PIGG catalogue has been used to measure the mass and luminosity
functions of groups and clusters. By combining the measured abundance
of clusters as a function of luminosity, with a typical
mass-to-light ratio, the cluster mass function is measured more
accurately than by using the dynamically-inferred masses
directly. After removing a bias due to measurement errors in the
abundance, using that found in mock catalogues constructed from
$\Lambda$CDM N-body simulations, the 2PIGG mass function implies that
$\sigma_8=0.25~\Omega_{\rm m}^{-0.92-4.5(\Omega_{\rm m}-0.22)^2}$ for
$0.18\le\Omega_{\rm m}\le0.50$. While the statistical uncertainty on
this value is around $10$ per cent, the systematic correction that
has been applied is closer to $20$ per cent. The uncertainty in this
correction, coupled with the systematic uncertainty in the
normalisation of the estimated cluster masses (the value of
$A$ in equation~\ref{mass}), currently limits the precision of this
determination of $\sigma_8$ to $\sim 20$ per cent. To decrease this
uncertainty significantly would require a more detailed understanding of how
the observed galaxies populate the underlying dark matter haloes.

The group luminosity function is in good agreement with previous
studies for groups with total $\bj$-band luminosity exceeding $\sim
4\times10^{10}\Lsol$. At lower luminosities, the abundance of 2PIGGs
is somewhat larger than has previously been found from smaller
samples. Matching this
abundance to that of haloes of a given mass in $\Lambda$CDM
simulations gives the mass-to-light ratio as a function of group
size. This indirect method combined with the direct estimates of Eke
\etal (2004b) at higher luminosities allows the recovery of the
variation of the 
group mass-to-light ratio over almost four orders of magnitude in
group luminosity. The resulting function has a minimum value of $\sim
100~hM_\odot/L_\odot$ at group luminosity 
$L_{b_{\rm J}}\sim5\times10^9\Lsol$.  
Finally, the group mass-to-light ratio was used to infer the halo
circular speed as a function of group luminosity. Comparison to
the observed Tully-Fisher relation for galaxies suggests that, 
at a given galaxy luminosity, the observed rotation speed is
similar to the typical halo circular speed. 

The dependence of the group mass-to-light ratio on group luminosity
provides a key test of various feedback processes invoked in models of
galaxy formation (Benson \etal 2000; Benson \etal 2003; Croton \etal
2005). The agreement
between the indirect determination at low luminosities presented in
this paper and the results of the semi-analytical $\Lambda$CDM model
assumed in the construction of the mock catalogues is encouraging. A
direct determination based, for example, on a deeper redshift survey
is desirable.

\section*{ACKNOWLEDGMENTS}

We would like to thank Christian Marinoni for his very helpful comments
concerning different wavebands. The Hubble Volume N-body simulation
was carried out by the Virgo Supercomputing Consortium using computers
based at the Computing Centre of the Max-Planck Society in Garching
and at the Edinburgh parallel Computing Centre.
VRE and CMB are Royal Society University Research Fellows. JFN
acknowledges support from the Alexander von Humboldt and Leverhulme
Foundations.

\end{document}